\documentclass[%
reprint,
superscriptaddress,
showpacs,
nofootinbib,
nobibnotes,
 amsmath,amssymb,
 aps,
prc,
floatfix,
]{revtex4-1}

\usepackage{graphicx}
\usepackage{dcolumn}
\newcolumntype{.}{D{.}{.}{-1}}
\usepackage{bm}
\usepackage{float,array,booktabs}

\begin{document}

\preprint{APS/123-QED}

\title{Second $\mathbf{ T = 3/2}$ state in $\mathbf{^9}$B and the isobaric multiplet mass equation}
\author{N.\,J.~Mukwevho }
\affiliation{Department of Physics and Astronomy, University of the Western Cape, P/B X17, Bellville 7535, South Africa}
\author{B.\,M.~Rebeiro}
\affiliation{Department of Physics and Astronomy, University of the Western Cape, P/B X17, Bellville 7535, South Africa}
\author{D.\,J.~Mar\'in-L\'ambarri}
\affiliation{Department of Physics and Astronomy, University of the Western Cape, P/B X17, Bellville 7535, South Africa}
\affiliation{iThemba LABS, P.O. Box 722, Somerset West 7129, South Africa}
\author{S.~Triambak}
\email{striambak@uwc.ac.za}
\affiliation{Department of Physics and Astronomy, University of the Western Cape, P/B X17, Bellville 7535, South Africa}
\author{P.~Adsley}
\affiliation{Department of Physics, University of Stellenbosch, Private Bag X1, 7602 Matieland, South Africa}
\affiliation{iThemba LABS, P.O. Box 722, Somerset West 7129, South Africa}
\author{N.\,Y.~Kheswa}
\affiliation{iThemba LABS, P.O. Box 722, Somerset West 7129, South Africa}
\author{R.~Neveling}
\affiliation{iThemba LABS, P.O. Box 722, Somerset West 7129, South Africa}
\author{L.~Pellegri}
\affiliation{School of Physics, University of the Witwatersrand, Johannesburg 2050, South Africa}
\affiliation{iThemba LABS, P.O. Box 722, Somerset West 7129, South Africa}
\author{V.~Pesudo}
\affiliation{Department of Physics and Astronomy, University of the Western Cape, P/B X17, Bellville 7535, South Africa}
\affiliation{iThemba LABS, P.O. Box 722, Somerset West 7129, South Africa}
\author{F.\,D.~Smit}
\affiliation{iThemba LABS, P.O. Box 722, Somerset West 7129, South Africa}
\author{E.\,H.~Akakpo}
\affiliation{Department of Physics and Astronomy, University of the Western Cape, P/B X17, Bellville 7535, South Africa}
\author{J.\,W.~Br\"ummer}
\affiliation{Department of Physics, University of Stellenbosch, Private Bag X1, 7602 Matieland, South Africa}
\author{S.~Jongile}
\affiliation{Department of Physics and Astronomy, University of the Western Cape, P/B X17, Bellville 7535, South Africa}
 \affiliation{Department of Physics, University of Zululand, Private Bag X1001, KwaDlangezwa 3886, South Africa}
 \author{M.~Kamil}
\affiliation{Department of Physics and Astronomy, University of the Western Cape, P/B X17, Bellville 7535, South Africa}
\author{P.\,Z.~Mabika}
\affiliation{Department of Physics and Astronomy, University of the Western Cape, P/B X17, Bellville 7535, South Africa}
 \affiliation{Department of Physics, University of Zululand, Private Bag X1001, KwaDlangezwa 3886, South Africa}
\author{F.~Nemulodi}
\affiliation{iThemba LABS, P.O. Box 722, Somerset West 7129, South Africa}
\author{J.\,N.~Orce}
\affiliation{Department of Physics and Astronomy, University of the Western Cape, P/B X17, Bellville 7535, South Africa}
\author{P.~Papka}
\affiliation{Department of Physics, University of Stellenbosch, Private Bag X1, 7602 Matieland, South Africa}
\affiliation{iThemba LABS, P.O. Box 722, Somerset West 7129, South Africa}
\author{G.\,F.~Steyn}
\affiliation{iThemba LABS, P.O. Box 722, Somerset West 7129, South Africa}
\author{W.~Yahia-Cherif}
\affiliation{Universit\'e des Sciences et de la Technologie Houari Boumediene (USTHB), Facult\'e de Physique, B.P. 32 El-Alia,16111 Bab Ezzouar, Algiers, Algeria}
\affiliation{iThemba LABS, P.O. Box 722, Somerset West 7129, South Africa}

%
%
%
 
\date{\today}
%
 \begin{abstract}
 Recent high-precision mass measurements and shell model calculations~[Phys. Rev. Lett. {\bf 108}, 212501 (2012)] have challenged a longstanding explanation for the requirement of a cubic isobaric multiplet mass equation for the lowest $A = 9$ isospin quartet. The conclusions relied upon the choice of the excitation energy for the second $T = 3/2$ state in $^9$B, which had two conflicting measurements prior to this work. We remeasured the energy of the state using the $^9{\rm Be}(^3{\rm He},t)$ reaction and significantly disagree with the most recent measurement. Our result supports the contention that continuum coupling in the most proton-rich member of the quartet is not the predominant reason for the large cubic term required for $A = 9$ nuclei.      
%
 \end{abstract}

\pacs{21.10.-k, 21.10.Hw, 21.10.Sfm, 25.55.-e, 27.20.+n}
\maketitle
If nuclear isospin $T$ were a conserved quantity, the members of an isobaric multiplet would have identical masses. However, it is well known~\cite{Benenson:79} that this degeneracy is broken by introducing two-body charge-dependent interactions. 
As a result of this isospin symmetry breaking, the masses of isobaric analog states (IAS) within a multiplet are related (to first order in perturbation theory) by the isobaric multiplet mass equation (IMME)~\cite{Wigner:57, Weinberg:59}
\begin{equation}  
 M(T_z) = a +bT_z + cT_z^2,
 \label{eq:imme} 
\end{equation} 
where each member of the multiplet is characterized by its isospin projection $T_z = (N - Z)/2$. Over the years, the widespread success~\mbox{\cite{Lam:13,Maccormick:14}} of the IMME as a local mass relation made it a useful tool to make predictions, particularly when direct measurements were difficult. For instance, it has been used to place bounds on scalar couplings in the weak interaction~\cite{Eric:99}, identify candidates for two-proton radioactivity~\cite{Dossat:05, Blank:08} and obtain thermonuclear reaction rates along the $rp$-process path~\cite{Richter:13,Ong:17}. In the recent past, the availability of Penning trap mass spectrometers at radioactive ion beam facilities, as well as the development of state-of-the-art computational techniques, such as the use of interactions based on chiral effective field theory~\cite{Gallant:14} or similarity renormalization group \textit{ab initio} calculations~\cite{Brodeur:17}, have enabled some of the most demanding tests of the IMME. 

Deviations from Eq.~\eqref{eq:imme} can arise if first-order perturbation theory is not sufficient to account for isospin non-conserving (INC) effects~\cite{Janecke}, or if many-body interactions are required~\cite{Bertsch:70}. The former becomes particularly relevant when the wave functions of the IAS differ significantly due to isospin mixing with nearby states of the same spin ($J$) and parity ($\pi$) or admixtures with unbound states (Thomas-Ehrman effects). In this context, the light mass $A = 7$, $A = 8$ and $A = 9$ multiplets present interesting case studies~\cite{Lam:13,Maccormick:14}. Many of the IAS in these nuclei are particle unbound, so that they contribute to sizable violations of the IMME~\cite{Janecke}. 

In order to further investigate INC effects in light nuclei, recent experiments at radioactive ion beam facilities have placed emphasis on the $A = 9,~T = 3/2$ quartet~\cite{Brodeur:12} and $A = 8,~T = 2$ quintet~\cite{CharityR:11,Charity:11}. Both these multiplets are known to exhibit significant departures from the quadratic form of the IMME, understood to be caused by isospin violating effects due to coupling with the particle continuum~\cite{Janecke,Bertsch:70}. Currently available data show that the former requires a cubic ($dT_z^3$) term, while the latter requires both cubic and quartic ($eT_z^4$) terms to obtain reasonable agreement with measured masses~\cite{Maccormick:14}.  

In this Rapid Communication we focus on the \mbox{$A = 9$} system. These light nuclei provide a fertile testing ground for recently developed tools in nuclear theory (such as \textit{ab initio} calculations) and have also been described using sophisticated cluster models~\cite{Oertzen:96,Arai:03,Vasilev:17,Zhao:18}. The latest compilation~\cite{Maccormick:14} shows that its first \mbox{$T = 3/2$} quartet requires a \mbox{$d = 6.7 \pm 1.5$~keV} cubic term for a satisfactory fit to the data. This value is consistent with the theoretical predictions ($d \approx 4$~keV) of Bertsch and Kahana, who used a combination of three-body second-order Coulomb and other charge-dependent nuclear interactions~\cite{Bertsch:70}. The enhanced $d$ coefficient for $A = 9$ is a natural consequence in their calculations, because of the weak binding of the last proton in $^9$C~\cite{Bertsch:70,Kashy:74}. 
 
Contrary to the above interpretation, in a recent publication by Brodeur~\textit{et al.}~\cite{Brodeur:12}, who performed high-precision mass measurements and shell model calculations, the large $d$ coefficient was attributed to isospin mixing in the $T_z = \pm 1/2$ members of the quartet. The purported $T = 1/2$ admixed states (of undetermined spin and parity) occur at excitation energies of $15100 \pm 50$ and $15290 \pm 40$ keV for $^9$Be and $^9$B respectively~\cite{Brodeur:12,nndc}. The shell model calculation of the cubic coefficient, using a PJT Hamiltonian~\cite{Julies:92,Brown:01} 
was found to be in excellent agreement with experiment~\cite{Brodeur:12}.  
Furthermore, if the IMME violation was indeed due to the wavefunction expansion of the particle-unbound $T = 3/2$ state in $^9$C, the deviation is expected to be significantly worse for the second $T = 3/2$ quartet, where both the $^9$B and $^9$C analog states have appreciable widths~\cite{nndc,Tilley:04}. However, on using the most precise available data~\cite{Brodeur:12} for the excited quartet, a cubic fit to the IMME yields a much smaller value of $d = 3.2 \pm 2.9$~keV. This is supported by the shell model calculations, which yield a cubic coefficient for the excited quartet that is consistent with zero. These results reinforce the isospin-mixing explanation for the IMME violation in the first quartet. 

In spite of the befitting agreement described above, a serious discrepancy arises if one takes into consideration the most recent determination~\cite{Charity:11} of the excitation energy of the second $T = 3/2$ state in $^{9}$B. On measuring the energies of the break-up particles from the $^9{\rm B} \to p + ^8{\rm Be}~(2\alpha)$ decay channel, the authors of Ref.~\cite{Charity:11} report the energy of the state to be $16990 \pm 30$~keV. This disagrees with the more precise determination of $E_x = 17076 \pm 4$~keV from an older $^{11}{\rm B}(p,t)$ measurement~\cite{Benenson:74} by around 3~standard deviations. If instead one uses the lower precision result from Ref.~\cite{Charity:11}, the cubic IMME fit yields a much larger $d = -40 \pm 15$~keV, which calls into question the explanation presented in Ref.~\cite{Brodeur:12}. This is a crucial aspect, as other similar experimental investigations in the past have been prone to misinterpretations on account of inaccurately known IAS excitation energies~\cite{
Gallant:14,Glassman:15,Herfurth:01,Pyle:02,Zhang:12,
Su:16}. To resolve the above issue, we remeasured the energy of the second \mbox{$J^\pi;T = 1/2^-;3/2$} state in $^9$B using the $^{9}{\rm Be}(^3{\rm He},t)$ reaction. 
\begin{figure}[t]
\includegraphics[scale=0.33]{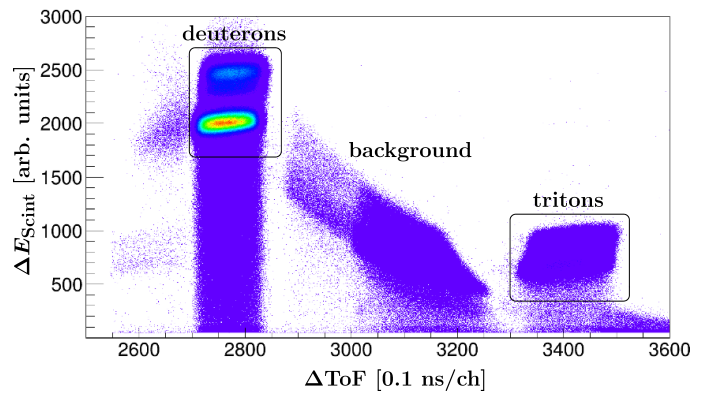}
\caption{\label{fig:pid}Particle identification spectrum using energy loss in the scintillator vs relative time of flight ($\Delta$ToF). $\Delta$ToF is the time difference between the cyclotron RF signal and the trigger from the scintillator. The background in the spectrum persists with an empty target frame and is most probably beam halo related.}
\end{figure}
\begin{figure}[t]
\includegraphics[scale=0.35]{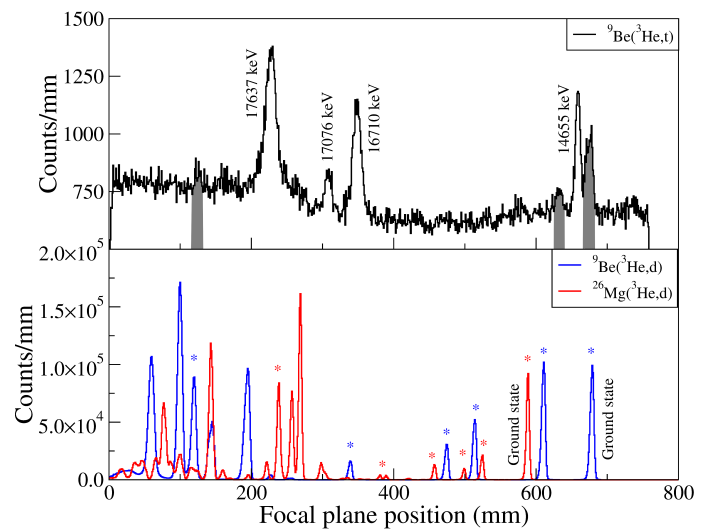}
\caption{\label{fig:spectra}Top panel: Triton spectrum from $^{9}{\rm Be}(^3{\rm He},t)^9{\rm B}$ in the 14-18~MeV excitation energy region. Peaks corresponding to previously known $^{9}$B energies are identified. The shaded regions indicate tentative new states in $^9$B that are not currently included in the evaluated nuclear structure (ENSDF) database~\cite{nndc}. These include a strongly populated doublet at approximately 14.6~MeV and two weaker triton peaks corresponding to excitation energies of 14.9~MeV and 18.3~MeV. Bottom panel: Calibration spectra from $^{9}{\rm Be}(^3{\rm He},d)^{10}{\rm B}$ and $^{26}{\rm Mg}(^3{\rm He},d)^{27}{\rm Al}$ reactions. The latter (red) is scaled by a factor of~15 for visualization purposes. Only peaks marked with asterisks were used for the energy calibrations.}
\end{figure}

In our experiment, a $50.61 \pm 0.05$~MeV pulsed, dispersion-matched $^3{\rm He}^{++}$ beam from the Separated Sector Cyclotron facility at iThemba LABS was bombarded on a 99.8\% pure, $4.4 \pm 0.2$~$\mu$m thick self-supporting $^9$Be target. The reaction products were momentum analyzed using the K600 magnetic spectrometer, operating in $0^\circ$~mode~\cite{Neveling:11}. The focal plane detectors of the spectrometer consisted of a multi-wire drift chamber (MWDC) followed by a 12.7~mm~thick plastic scintillator. The MWDC determined the horizontal and vertical positions of the light charged ejectiles crossing the focal plane, while the plastic scintillator was used for particle identification (PID) purposes and as a trigger detector. 

A sample PID spectrum is shown in Fig.~\ref{fig:pid}, which highlights a clear discrimination between tritons and deuterons from $^{3}{\rm He}$ induced reactions on the $^{9}$Be target. 
The upper panel of Fig.~\ref{fig:spectra} shows the focal plane spectrum obtained using the appropriate spectrometer field settings and further gating on the triton group. The lower panel in Fig.~\ref{fig:spectra} shows deuteron spectra obtained using the same field settings, from both the $^9$Be target and an additional $3.9\pm 0.1~\mu$m thick $^{26}$Mg target, which was isotopically enriched to 99.4\%. Both these spectra were used for energy calibration purposes. The thicknesses of the targets were determined using a $^{226}$Ra~$\alpha$~source and an iterative algorithm that used infinitesimal slices of target thickness and SRIM~\cite{srim,srim2} to determine initial stopping powers for the unattenuated $\alpha$ energies\footnote{In all our analysis we conservatively assume 10\% relative uncertainties in the stopping powers obtained from SRIM. It is assumed that transverse component contributions to the energy losses play an insignificant role.}. 

The energy resolution of the spectrometer was determined 
to be comparable to (or worse than) the intrinsic widths of the states highlighted in Fig.~\ref{fig:spectra}. 
This feature allowed us to fit the triton peaks using a simple function comprised of a Gaussian on a flat background. A similar maximum likelihood procedure was also used to fit the deuteron spectra. However, these fits yielded better $\chi^2$ values on using a lineshape function that was the convolution of a Gaussian with a low-energy exponential tail. This is not surprising, considering that the two reactions have different kinematics and the spectrometer was optimized for the $(^{3}{\rm He},t)$ reaction. 

Once the peak centroids were obtained, a relativistic kinematics code was used to calibrate the deuteron momenta along the focal plane of the spectrometer and further determine $^9$B excitation energies. We briefly describe our analysis procedure below. 

Two important factors in the analysis were the location of the reaction(s) in the target(s) along the beam axis, and the corrections arising from energy losses. In order to take these into consideration, we first generated a momentum distribution for the deuterons from both calibration reactions, with Monte Carlo simulations that assumed randomly distributed reaction locations within the target from a uniform probability density function. The reduced $^3$He energy at a given randomized location was obtained from a numerical integration~
and interpolated values of energy losses from SRIM~\cite{srim,srim2}. The reaction kinematics was then used to calculate deuteron momenta corresponding to random values of reaction location. Following this, similar energy-loss calculations were carried out to obtain the final momenta of the deuterons exiting the target. Histogramming these values for the simulated events showed that the outgoing deuteron momenta were also uniformly distributed, between values $p_{\rm min}$ and $p_{\rm max}$ corresponding to reactions on the back and the face of the target respectively. The average momenta ($\bar{p}_d$) of the deuterons were then simply $(p_{\rm min}+p_{\rm max})/2$, given the flat nature of the deuteron momentum distribution~\cite{CowanBook}. The $\bar{p}_d$ values corresponding to well-resolved excited states in $^{10}$B and $^{27}$Al (highlighted in Fig.~\ref{fig:spectra}) were used to calibrate the momenta of the 
ejectiles detected at the focal plane of the spectrometer. This was performed using a quadratic regression with respect to the peak centroids $\mu(i)$,
\begin{equation}
 \bar{p}_d(i) = a_0 + a_1 \mu(i) + a_2 \mu(i)^2.
 \label{Eqn_FP_momentum}
\end{equation}

The $^{9}$B excitation energies of interest were finally calculated from the triton momenta evaluated using the parameters of the above fit, triton energy loss corrections\footnote{A reconstruction of the reaction locations $x(i)$ for average ejectile momenta $\bar{p}(i)$ showed that the $x(i)$'s correspond the center of the targets.} and the $^{9}{\rm Be}(^3{\rm He},t)$ reaction kinematics.
\begin{table}[t]
\caption{\label{tab:results}$^9$B excitation energies obtained from this experiment using both calibration reactions.
Our adopted value is from the weighted mean of the results from the two calibrations, while retaining the (smaller) statistical uncertainties. These uncertainties are added in quadrature with our conservative estimates of systematic uncertainties from the sources listed in Table.~\ref{tab:errors}.  
}
\begin{ruledtabular}
\begin{tabular}{cccc}
&\multicolumn{2}{c}{Measured energies [keV]$^a$}&\\
\cline{2-3}\\[-0.9em]
\multicolumn{1}{c}{Previous work$^b$} &\multicolumn{1}{c}{$^{26}{\rm Mg}(^3{\rm He},d)^c$}&\multicolumn{1}{c}{$^{9}{\rm Be}(^3{\rm He},d)$}&\multicolumn{1}{c}{This work}$^d$\\
\multicolumn{1}{c}{[keV]}& \multicolumn{1}{c}{calibration}& \multicolumn{1}{c}{calibration}&\multicolumn{1}{c}{[keV]}\\
\colrule\\[-0.9em]
...                &$14538 \pm 2$ &$14538 \pm 2$ &$14538 \pm 19$\\
...                &$14582 \pm 4$ &$14583 \pm 4$ &$14582 \pm 19$\\
$14655.0 \pm 2.5$  &$14663 \pm 1$ &$14665 \pm 1$ &$14664 \pm 19$\\
...                &$14842 \pm 3$ &$14847 \pm 3$ &$14845 \pm 19$\\
$16710 \pm 100$    &$16790 \pm 1$ &$16795 \pm 1$ &$16792 \pm 19$\\
$16990 \pm 30^{e}$ &              &              &\\
$17076 \pm 4^{f}$  &$17071 \pm 3$ &$17074 \pm 3$ &$17073 \pm 19$\\
$17637 \pm 10$     &...           &$17627 \pm 1$ &$17627 \pm 19$\\
...                &...           &$18329 \pm 5$ &$18329 \pm 20$      
\end{tabular}
\end{ruledtabular}\\
\begin{flushleft}
$^a$ Only statistical uncertainties are listed in these columns.\\
$^b$ $E_x$~from Refs.~\cite{nndc,Tilley:04}.\\
$^c$ We do not use the $^{26}{\rm Mg}(^3{\rm He},d)$ reaction to calibrate the two highest energy peaks. This is because they require significant extrapolations, as evident in Fig.~\ref{fig:spectra}.\\ 
$^d$ With systematic uncertainties added in quadrature.\\
$^e$ Second $T = 3/2$ state from Ref.~\cite{Charity:11}.\\
$^f$ Second $T = 3/2$ state from Ref.~\cite{Benenson:74}.
\end{flushleft}
\end{table}
\begin{table}[t]
\begin{flushleft}
\caption{Relative contributions of systematic uncertainties in determining the excitation energy of the second $T = 3/2$ state in $^9$B.}
\label{tab:errors}
\begin{ruledtabular}
\begin{tabular}{l.}
Source & \multicolumn{1}{c}{$\Delta E_x/E_x$~[\%]} \\
\colrule\\[-0.95em]
Ground state masses& 0.004   \\
Beam energy$^a$& 0.092 \\
Target thickness& 0.01 \\
Ejectile momenta~($\bar{p}_d$) used for calibration$^b$&  0.05 \\
Stopping powers &0.008 \\
\colrule\\[-0.9em]
Total & 0.11 
\end{tabular}
\end{ruledtabular}\\[0.2em]
$^a$ An overly conservative estimate of the uncertainty in the beam energy is $\pm 50$~keV. This arises from a determination of the bending radius in the analyzing magnet located upstream the K600 spectrometer. \\   
$^b$ The variance of a uniform distribution that is bounded between values $\alpha$ and $\beta$ is $\frac{1}{12}
(\beta - \alpha)^2$~\cite{CowanBook}. 
\end{flushleft}
\end{table}

Table~\ref{tab:results} lists the energies of relevant $^9$B states that were extracted using the procedure described above. As accentuated by the shaded regions in  Fig.~\ref{fig:spectra}, we identify four new tentative states in $^9$B (also listed in Table~\ref{tab:results}) that are not included in the latest compilation for $A = 9$ nuclei~\cite{Tilley:04}. The possibility of these peaks arising from typical contaminants such as $^{12}$C and $^{16}$O can be easily ruled out on account of the large differences in reaction $Q$ values. Nevertheless, a more comprehensive analysis of these states is recommended. 

Our determined excitation energy for the second \mbox{$T = 3/2$} state in $^9$B is in almost exact agreement with the older $^{11}{\rm B}(p,t)$ measurement~\cite{Benenson:74}, while being significantly different from the most recent measurement of Charity \textit{et al.}~\cite{Charity:11}.  We infer that despite the comprehensive data analysis procedure described by the latter, which includes an irrefutable confirmation of the spin and parity of the $1/2^-$ state, it is quite likely that unaccounted systematic effects undermined the accuracy of their excitation energy measurement. On the other hand, the $^{11}{\rm B}(p,t)$ measurement was performed with a split-pole magnetic spectrograph and the energy of the state was quoted with a relative precision of $\approx 0.02$\%. Notwithstanding the sparse description of systematic effects in this particular work, it is reassuring to note that the same authors provide an adequate description of their beam energy calibration using a momentum matching technique~\cite{
Trentelman} and consideration of target-thickness effects in other similar measurements~\cite{Trentelman_prl,Kashy:75,Kashy:74}.      

In light of the above, the result of our measurement gives credence to the isospin-mixing explanation put forth by Brodeur~{\it et al.}~\cite{Brodeur:12}. It arguably rules out the longstanding hypothesis that the requirement of a large cubic term to the IMME for the $A = 9$,~$T = 3/2$ quartet is mainly due to the extended orbit of the least-bound proton in $^{9}$C~\cite{Bertsch:70,Kashy:74}. Contrary to what would be expected in such a scenario, a cubic IMME fit to the second $T = 3/2$ quartet using the weighted mean of the three measured excitation energy values yields $d = 2.4 \pm 2.9$~keV, which is consistent with zero.

In summary, we have measured the excitation energies of several high-lying states in $^9$B using $(^3{\rm He},t)$ and $(^3{\rm He},d)$ reactions, while placing a rigorous emphasis on energy loss corrections and other systematic uncertainties in our analysis. Our determined excitation energy for the second $T = 3/2$ state in $^9$B (with $\Delta E/E \approx 0.1\%$) disagrees with the latest preceding measurement~\cite{Charity:11}. Consequently, our result provides ample evidence to exclude a previously well-established hypothesis~\cite{Bertsch:70,Kashy:74}, that invoked continuum-coupled effects to explain the cubic IMME for the lowest $A = 9$ isospin quartet. As by products of this work, we identify four previously unknown states that require further investigation. We also obtain an improved determination of the excitation energy for the provisional $\mbox{$J^\pi;T = 5/2^+;1/2$}$ state at 16.8~MeV~\cite{Tilley:04}.    

We are thankful to Alejandro~Garc\'ia and Paul Garrett for fruitful discussions. NJM acknowledges financial support from the National Research Foundation of South Africa (NRF) during the course of his MSc research. RN acknowledges support from the NRF through Grant No.~85509. The role of the iThemba LABS accelerator group in ensuring smooth beam delivery and operation is gratefully acknowledged. The enriched $^{26}$Mg target material was supplied by Oak Ridge National Laboratory. %
%

\bibliography{9b_imme}

\end{document}